\documentclass[12pt]{article}
\usepackage{times}
\usepackage{geometry}
\usepackage{graphicx}
\usepackage[utf8]{inputenc}
\usepackage{enumitem,amssymb}
\usepackage{ragged2e}
\usepackage{natbib}

\newlist{thematic}{itemize}{8}
\setlist[thematic]{label=$\square$}
\usepackage{pifont}
\begin{document}
\large
%AGN Coronae White Paper \linebreak

%Replace with Your Title \linebreak

\title{\textbf{Astro 2020 Science White Paper}: \\ Probing the Physical Properties of the Corona in Accreting Black Holes} 

% MB: The WP seems undecided on whether it talks about black holes more generally or just AGN. Except for a casual mention of BHBs here and there, not much is said about them. So, it's unclear if the title should be AGN-specific or broader.
% NK: We could keep it broad by referring to accreting black holes rather than AGN, so that the title doesn't sound too specific. Maybe something like "Probing the Physical Properties of the Corona in Accreting Black Holes"?
% MB: Ok, the title is more general, which works fine, but the paper is still essentially about AGN coronae. Have in mind that some people are alergic to titles that start with "probing".

\author{Nikita Kamraj$^1$, Andrew Fabian$^2$, Anne Lohfink$^3$, Mislav Balokovi\'{c}$^4$, \\
Claudio Ricci$^{5,6}$, Kristin Madsen$^1$}

\date{}
\maketitle

\vspace{-1cm}
\begin{flushleft}
\small
$^1$ Cahill Center for Astronomy and Astrophysics, California Institute of Technology, Pasadena, CA 91125, USA \\
$^2$ Institute of Astronomy, Madingley Road, Cambridge, CB3 0HA \\
$^3$ Department of Physics, Montana State University, 211 Montana Hall, Bozeman, MT 59717, USA \\
$^4$ Center for Astrophysics $|$ Harvard \&  Smithsonian, 60 Garden Street, Cambridge, MA 02140, USA \\
$^5$ N\'{u}cleo de Astronom\'{i}a de la Facultad de Ingenier\'{i}a, Universidad Diego Portales, Av. Ej\'{e}rcito Libertador 441, Santiago, Chile \\
$^6$ Kavli Institute for Astronomy and Astrophysics, Peking University, Beijing 100871, China \\
%$^7$ Chinese Academy of Sciences South America Center for Astronomy, Camino El Observatorio 1515, Las Condes, Santiago, Chile
%\noindent 
\vspace{5pt}
\textbf{Contact:} nkamraj@caltech.edu
\vspace{1cm}

\normalsize \textbf{Thematic Areas:} Formation and Evolution of Compact Objects, Galaxy Evolution 

\vspace{5pt}

\end{flushleft}

\begin{flushleft}
\normalsize
\justify
\noindent
\textbf{Abstract:} The corona is a key component of most luminous accreting black holes, carrying 5 -- 30~\% of the power and in non-jetted Active Galactic Nuclei (AGN), creating all the X-ray emission above $\simeq 1-2$\,keV. Its emission illuminates the inner accretion disc, creating the atomic line-rich reflection spectrum used to diagnose and map the accretion flow and measure black hole spin. The corona is likely powered magnetically by the strong differential rotation of the accretion disc and it may be intimately related to relativistic jets. Recent work shows that many black hole coronae may be dominated by electron-positron pairs produced by photon-photon collisions in the compact coronal environment. Despite the corona being an integral component of AGN and black hole binary systems, much is still unknown about the nature of the corona, such as its geometry, location, and the physical mechanisms powering the emission. In this white paper we explore our current understanding of coronal properties, such as its temperature, obtained from measurements with existing hard X-ray telescopes such as \emph{NuSTAR}, and discuss important questions to be addressed in the coming decade surrounding the nature of the corona. Hard X-ray observations will continue to dispel the mystery of coronae and open up this part of the quasar engine to full understanding.
\end{flushleft}

%
% NOTES:
% White paper text can be max of five pages (p. 2--6) including figures.
% Due date is March 11, 2019.
% 

\pagebreak
\vspace{-8pt}
\section{Introduction to Coronae}
\vspace{-8pt}
The corona is an important component of most luminous accreting black holes, creating the variable, hard ($>2$\,keV) X-ray emission that is commonly observed in Active Galactic Nuclei (AGN). The corona is named after the Solar corona but its luminosity in AGN is more than $10^{15}$ times greater, despite its physical extent being similar in black holes systems of $10^{6} - 10^{7}M_{\odot}$. Its properties scale with mass above and below that range. It is generally considered that the corona is composed of a hot cloud of plasma located close to the accretion disc \citep{corona-haardt}. Compton upscattering of UV/optical photons from the accretion disc by electrons in the corona produces the intrinsic X-ray continuum \citep{rybicki-lightman}.
% MB: The two sentences below don't seem necessary; commenting out.
% Studies of the corona enable us to understand how quasars work, in the full sense. The accretion process and in particular the accretion disc generates the power but the corona mediates that power in a way that is useful to us.

The coronal X-ray emission illuminates the accretion disc, which gives rise to the reflection spectrum containing atomic lines, most notably of iron, which are used to diagnose and map the accretion flow. Doppler shifts and gravitational redshift enable the inner edge of the disc to be located and, assuming that it is the Innermost Stable Circular Orbit (ISCO), mean that the spin of the black hole can be measured \citep{reynolds-2014}. 
%AL: Maybe mention the power output again here as a motivation to why the corona is important? The reflection things kind of fall from the sky and 'appear' to be unconnected to our WP.
%CR: add some references here. NK: Added BH spin ref.
% MB: Merging these two pieces above into a single paragraph; they seem to fit fine.
Reverberation mapping studies and the observed rapid X-ray variability from many AGN indicate that the corona is physically small, of the order 3--10 gravitational radii (r$_{g} = GM/c^{2}$) in radius and height above the black hole \citep[e.g.,][]{Emmanoulopoulos-2014,uttley-2014}. Observations of microlensing in quasars also indicate compact X-ray emitting regions \citep{chartas-lensing}. 
% MB: This may be a good spot to include some BHB references (also for spin in the part above).

\vspace{-8pt}
\section{Compactness and the $\Theta-l$ Plane}
\vspace{-8pt}
The corona is not only physically small, but also compact in the radiative sense, meaning that the ratio of luminosity to radius, $L/R$,  is large. The compactness is usually described by the dimensionless parameter $l = \frac{L}{R} \frac{\sigma_{T}}{m_{e}c^{3}}$, where $\sigma_{T}$ is the Thompson scattering cross section and $m_{e}$ the electron mass \citep{guilbert-1983}. The density of high energy photons can be high enough that photon-photon collisions can lead to electron-positron pair production. Increased energy fed to the corona then has the effect of producing more particles to share the energy, limiting any rise in coronal temperature and preventing pair production from quickly becoming a runaway process and exceed annihilation. It thereby acts as an $l$-dependent thermostat  \citep{svensson-1984,Zdziarski-1985,stern-1995}. The coronal temperature can also be characterised by the dimensionless parameter $\Theta = k_{B}T/m_{e}c^{2}$, where $k_{B}$ is the Boltzmann constant. In determining coronal temperatures from X-ray spectra, it is usually assumed that $k_{B}T = E_{cut}/2$, where $E_{cut}$ is the high-energy cutoff in the spectrum \citep{petrucci-2001}.

Within the corona, two-body collisions are the simplest heating and thermalization mechanism. It is commonly assumed that the corona is magnetically powered \citep{merloni-2001} by the strong differential rotation of the accretion disc. Radiation pressure may cause a pair wind if there are open field lines from the corona. Twisted field lines extending along the black hole spin axis may accelerate a jet of pairs from the corona. The pair thermostat behaviour defines a line in the $\Theta-l$ plane to the right of which a static physical corona cannot exist. The precise location of the pair line depends on the geometry of the corona and nature of the soft photon field. Figure 1 illustrates the location of some pair lines for different geometries, such as a slab and hemispherical corona, along with particle coupling lines.

%\begin{figure}[h]
%    \centering
%    \includegraphics[width=0.6\textwidth]{theta-l-lines-plot.pdf}
%    \caption{The $\theta-l$ plane, showing boundaries for runaway pair production, e-e coupling, e-p coupling, and the line where Compton cooling dominates over bremsstrahlung cooling ($t_{B} = t_{C}$).}
%    \label{fig:theta-l-pair-lines}
%end{figure}
% NK: Commenting out plot as it is reproduced in section 3

Observations of AGN X-ray spectra show that most AGN do indeed lie to the left of the $\Theta-l$ line \citep{fabian-2015}. More recently, a significant number of AGN exhibit coronal temperatures that are much lower than expected for a fully thermal pair plasma \citep[e.g.,][]{grs-low-ecut,kara-2017,yanjun-2017}. One possible means of producing a low temperature corona is by considering the corona to be hybridized, composed of a mixture of thermal and non-thermal particles \citep[e.g.,][]{Zdziarski-1993,corona-heating-paper,fabian-2017}. Only a small fraction of non-thermal electrons with energies above 1 MeV would be needed to result in runaway pair production. The cooled pairs share the energy, resulting in a reduction of the coronal temperature \citep{fabian-2017}.

\vspace{-8pt}
\section{Current Modeling and Measurements}
\vspace{-8pt}
\begin{figure*}
\centering
\includegraphics[width=0.4\textwidth]{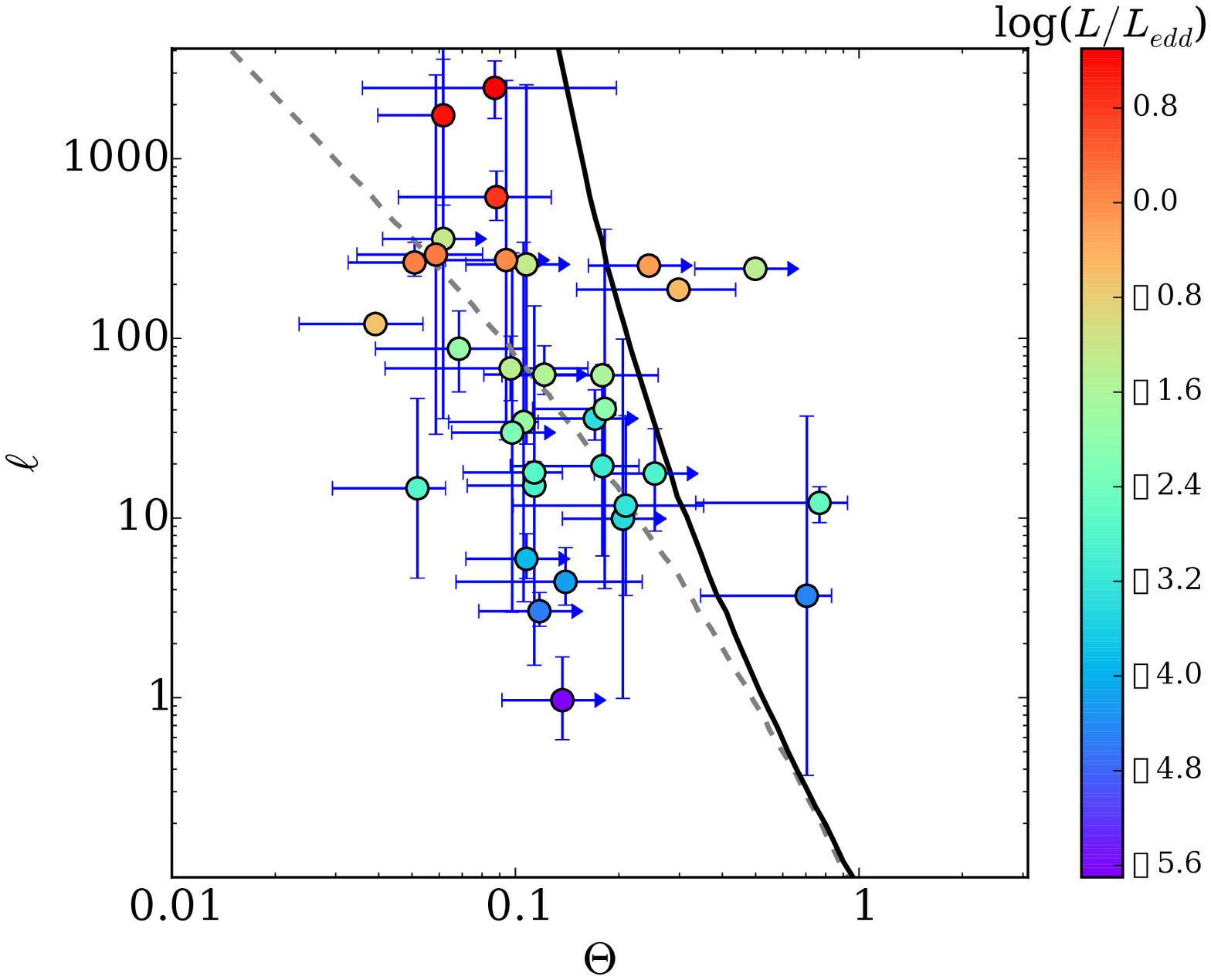}
\includegraphics[width=0.5\textwidth]{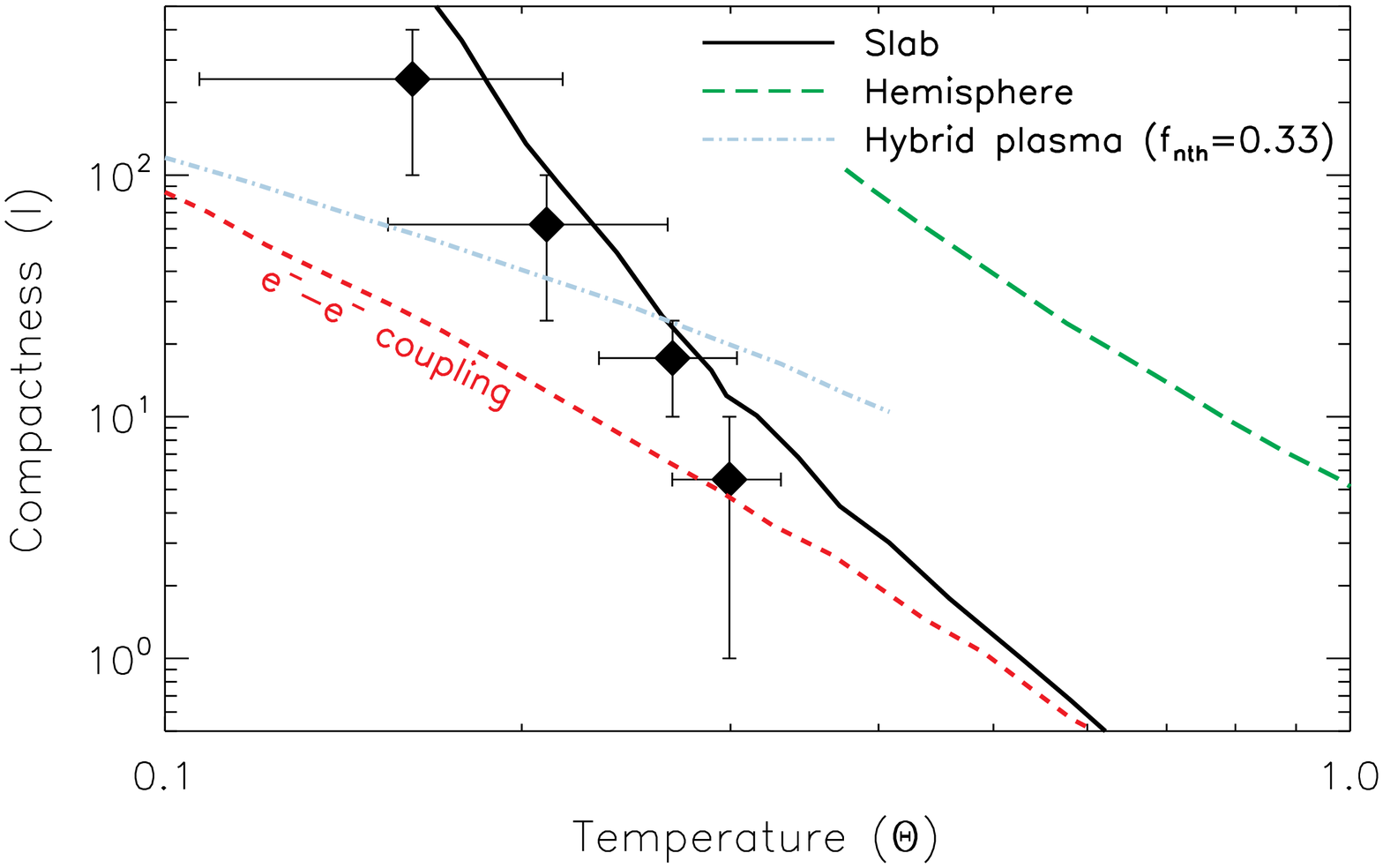}
\caption{{\it Left panel:} Compactness–temperature diagram for sources with a cutoff energy inferred by {\it NuSTAR} observations, color-coded based on their Eddington ratio (updated from \citealp{ricci-2018}).
The continuous line represents the pair line for a slab corona. 
{\it Right panel:} Same as the left-handed plot, with median values of the compactness and temperature for the unobscured AGN from the {\it Swift}/BAT sample \citep{ricci-2018}. The red dashed curves show the boundary to the region dominated by electron–electron coupling, while the dashed green and continuous black curves represent the runaway pair production limits for a a hemisphere and a slab corona, respectively. The runaway pair production limit obtained considering a hybrid plasma with 33 \% of non-thermal electrons is shown by the cyan dot–dot–dashed curve. }
\label{fig:thetaellnustar}
\end{figure*}

Current spectral modeling of the Comptonization spectrum can be divided into two categories: phenomenological modeling with a cut-off power law and the description with physical models. These physical models vary in their assumptions for the spectral shape of seed photons, geometry of the Comptonizing region and amount of radiation physics included. Popular Comptonization models such as \texttt{compTT} (\citealp{titarchuk-1994}), which allows the user to choose different simple geometries, have been used extensively. More recent, physically more self-consistent calculations such as \texttt{eqpair} \citep{coppi-1999}, \texttt{BELM} \citep{belm-2008}, and \texttt{MoCa} \citep{tamborra+2018-MoCa}, include a variety of geometries, hybrid coronae, and polarization.
% MB: I modified the sentence above to stress that they are more recent, added a new one and mentioned polarization. I know this section is on current modeling, but I also think that we must show that in terms of theory the field is not stale.
All these models describe the current data equally well in most cases.
% MB: This last sentence invites doubt that models can be distinguished (ever). If others think it should stay, since it's not incorrect, I suggest to add "current" or "existing" in front of "data".
% NK: added "current"

The high sensitivity of \textit{NuSTAR} above 10\,keV has transformed studies of the X-ray Comptonization spectrum of AGN, allowing measurements of the high-energy cut-off from single epoch observations in relatively large samples of nearby AGN for the first time. Previous studies were only focused on the brightest nearby AGN (e.g., \citealp{Nicastro:2000rw,Lubinski:2016la}).
This has resulted in numerous studies exploring the Comptonization spectrum of single objects in great detail (e.g., \citealp{ballantyne-2014,Matt:2014nx,mislav-2015,Tortosa:2017qe}). Some of these studies have been able to not only constrain the coronal temperature by detecting and measuring the high-energy roll-over of the spectrum and but also its variability \citep{ballantyne-2014,zoghbi-2017,ursini+2018-3c382}.
Although these variations of the cut-off energy are established, it is unclear how they can be understood in the context of the pair coronae model. Do magnetic instabilities cause the coronal heating to be variable or are geometrical changes behind the observed temperature changes?  
% MB: Rather than just saying that this is not understood, we should offer one or two interesting-sounding and currently unverifiable options that clearly require this research to continue, which is presumably why we are writing this white paper.
%AL: I am adding a sentences here, if we need space we can delete them again.

The more-readily available high quality hard X-ray data has also allowed the study of larger samples of sources. Most notable is the \textit{Swift}-BAT sample, comprised of all those AGN detected by the all-sky 14--195\,keV BAT survey (e.g., \citealp{Baumgartner:2013ee,Oh:2018tg}). Analysis of the spectra obtained with BAT over the course of the \textit{Neil Gehrels Swift} mission has shown that a high-energy cutoff is ubiquitous in the X-ray spectra of AGN, and that the median cutoff energy of local unobscured AGN is $E_{\rm C}=210\pm36$\,keV \citep{Ricci:2017sp}.
These studies have also found correlations of cut-off energy with other physical parameters of the accreting black hole. In particular, the cut-off energy has been shown to be connected to the Eddington ratio, $\lambda_{\rm Edd}$ \citep{ricci-2018}: 
AGN with $\lambda_{\rm Edd} > 0.1$ are found to show a lower median cut-off energy ($E_{\rm C} =
160 \pm 41$\,keV) than those with $\lambda_{\rm Edd} \leq 0.1$ ($E_{\rm C} = 370 \pm 51$\,keV). This result is in good agreement with the idea that coronae which are radiatively compact are also cooler, since they are rarely found in the region in the temperature-compactness parameter space dominated by runaway pair production.
Studies aimed at better understanding the relationship between coronal properties and the physical characteristics of accreting supermassive black holes exploiting \textit{NuSTAR} \citep{nikita-2018,tortosa-2018} are currently ongoing.

%Nustar fails by itself to characterize the shape of the high energy cut-off.
\vspace{-8pt}
\section{Future Goals}
\vspace{-8pt}
% LIST OF POSSIBLE TOPICS TO COVER:
% Study variability of Corona systematically - does kT respond to ell? Test thermostat. 
% Accumulate large samples of objects with good measurements, so can divide into separate classes. AGN vs BHB. Any mass dependence? What is dependence on Eddington Fraction.
% Detailed modelling with more complex models able to deal with different geometries.
% How does the corona vary? Is it associated with QPOs?
% Is the pair annihilation line observable? Tentative INTEGRAL results on NGC4151 (Lubinski+10, Siegert 16). 
% Plasma is probably hybrid. What fraction is nonthermal? Why? Can we see the nonthermal tail in hard X-ray spectra?
% How are pairs retained or are many blown out? How strong can pair winds be? What is the connection to jets? Do pairs from the corona become the pairs in the jet? Is any coronal power from the BZ mechanism? If so how much?

% From ACF:
Detailed studies of black hole coronae have only just begun. There is considerable scope to determine the geometry, i.e. the size, shape and location, of the corona in individual sources using relativistic reflection and reverberation of the coronal emission. Low frequency lag signals are likely due to small spectral changes in the coronal emission. The thermal/non-thermal nature of the coronal plasma can be found from observation of any non-thermal tail to the emission and the pair content from detection of the 511 keV annihilation line \citep{lubinski-2010,siegert-2016}. The likely hybrid nature of the corona raises the question of the fraction of non-thermal particles present. In order to investigate the presence of features such as hard non-thermal tails in X-ray spectra and robustly test hybrid plasma models, next-generation hard X-ray observatories with high sensitivity at energies beyond 100 keV will be essential.    
% NK: Added lines on hybrid coronae and removed "Pair winds are an interesting possibility given the strong radiation pressure present in such luminous objects" since it was mentioned in section 2 already.

% NK: Some of this is repeated in the paragraph on relativistic jet models - commenting out here and expanding later paragraph. "How the corona connects with a jet can be probed using high frequency radio emission which may correlate with the X-rays. In the case of stellar mass black hole systems,  rapidly variable near IR emission can be correlated \citep[e.g.,][]{gandhi-2010,gandhi-2016}. Theoretical modeling is required to explain exactly how the corona is powered. Are reconnection, shocks, or twisted magnetic fields responsible? Is the  heating intermittent and due to rapidly changing regions yet smaller  than currently envisaged?"

% Multiwavelength picture - other observational probes of coronae in other wavelength bands
Coronae are typically unobservable outside of the X-ray band. However, if probed at sufficiently high resolution ($<\!1$'') and high frequency ($>\!30$\,GHz), radio emission in radio-quiet (RQ) AGN may be traced to the X-ray-emitting corona \citep{inoue+doi-2014}. Beyond the tentative scaling between luminosities, $L_{\rm X}/L_{\rm R}\approx 10^{4}-10^{5}$ (\citealt{smith+2016}, \citealt{behar+2018}), sensitive high-resolution and high-frequency observations of nearby RQ AGN with ALMA, VLA, and VLBA are expected to reveal spectral shapes of the radio emission from hybrid coronae (e.g., \citealt{inoue+doi-2018}, \citealt{panessa+2019}).

% What is size, shape and location of coronae? Evidence from coronal emission as well as reflection and reverberation
% Coronae in jetted objects. Compare with non-jetted. How does it respond to jet formation?
Some relativistic jet models invoke the corona as the base of a jet \citep{merloni+fabian-2002}.
% While some AGN develop powerful jets that propagate to kpc and Mpc distances, in RQ AGN the subsequent acceleration and collimation may be missing.
Future studies that tie together measurements of the coronal temperature from hard X-ray spectroscopy with constraints on coronal size and location from reverberation measurements in the soft X-ray band  (e.g., \citealt{wilkins+2015}) and properties of relativistic jets in the radio band (e.g., \citealt{king+2017}) will pave the way to understanding the physical relationship between AGN coronae and jets. High-throughput observatories with broadband X-ray coverage such as {\em STROBE-X} \citep{ray+2019-strobex} or {\em HEX-P} \citep{madsen-2018},
% especially if combined with {\em Athena} or {\em Lynx},
% MB: I leave the above commented out because I'm not sure it's necessary; it makes the sentence a bit confusing, and I also don't have the references. 
would enable both of these key X-ray measurements for large samples of nearby AGN.

% Polarization of coronal emission
% How is corona powered exactly? Reconnection? Electric fields from twisted magnetic fields? Shocks? Is the heating intermittent (probably is)? Are powered regions much smaller than assumed so far?
% What is magnetic field geometry?
% What is connection with the Lusso-Risaliti results? (alpha$_{ox}$ - Used to perhaps measure H$_{0}$).
As X-ray polarimetry becomes available in the early 2020-ies with IXPE \citep{weisskopf-2018}, studies of polarimetric properties of coronae in both X-ray and radio promise to reveal the geometry of the hot plasma, and possibly properties of the magnetic field that permeates it.
% In analogy with stellar coronae,
AGN coronae may be powered by magnetic reconnection \citep{di-matteo-1998}, which would imply that heating is intermittent, and emission is variable. In the case of stellar mass black hole systems, rapidly variable near IR emission can be correlated with X-ray variability \citep[e.g.,][]{gandhi-2010,gandhi-2016}.
% MB: In the sentence above it's a bit unclear, correlated with what? NK: X-ray variability; sentence modified
Future multi-wavelength monitoring could reveal the variability of coronal emission in relation to variability of other AGN components, like the accretion disk, to which it is inextricably connected (e.g., \citealt{mehdipour+2016-ngc5548}).
% NK: Was there a particular paper you were thinking of? Not sure how much time I have to look at this in depth.
% MB: I'll see if I can add it; if not, this paragraph is still ok. Note: added.
With sufficient understanding of the disc-corona system and its non-linear scaling with luminosity, optical (rest-frame UV) data readily available from large-area surveys (e.g., LSST), and X-ray observations of unobscured quasars at cosmological distances may soon start to provide competitive constraints on cosmological parameters \citep{lusso+risaliti-2017}.

% In the era of high-cadence large-area surveys in the optical (LSST) and infrared (SPHEREx), it will be increasingly important to understand the dynamics of the evolving disk-corona systems now broadly classed as "changing-look AGN". They may be in some way related to state changes in BHBs
% MB: There is a brand new paper by Ruan et al. posted days ago that argues in this direction, but I guess there's not enough space to go into it without saving space somewhere else. This is the paper: https://arxiv.org/abs/1903.02553

%NK: Adding summary sentences below
% AL: Do we really not want to set the future telescopes up better? 
% NK: Do you have any suggested modifications? We could mention more specific instruments but that may take away from the science focus of this WP
% MB: See the penultimate section in Kristen's HEX-P paper -- it contains some nice and strong statements that would help to end this WP on a high note: https://pcos.gsfc.nasa.gov/physpag/probe/HEXP_2016.pdf There seems to be space for one or two more lines of text (unless it's added elsewhere).
%NK: Added a sentence based on some statements from Kristin's WP.
In summary, while the corona is a fundamental component of luminous black hole systems, responsible for powering the X-ray emission in AGN, much is still unknown about its nature, and even origin. Current observations of AGN with hard X-ray satellites such as \emph{NuSTAR} have allowed significant progress to be made in the field, through measurements of the high energy cutoff and in-depth studies of the Comptonization spectra of AGN for the first time. However, coronal measurements made using \emph{NuSTAR} are based on subtle downturns in continuum spectra near $\sim$ 50 keV, which are used to infer cutoff energies of $\sim$ 100 keV or higher. Future observations with next-generation hard X-ray telescopes such as \emph{HEX-P}, % \citep{madsen-2018} MB: I moved the first mention of HEX-P and the reference two paragraphs back.
combined with multi-wavelength monitoring will be of critical importance for deepening our understanding of the corona and answering the questions briefly discussed in this white paper within the coming decade.

\pagebreak
%\textbf{References}

% bibliography:

\bibliographystyle{aasjournal} % .bst file
\bibliography{corona_wp_refs} % .bib file

\end{document}